\documentstyle[12pt]{article}
\makeatletter
\begin{document}

\topmargin 0pt

\oddsidemargin -3.5mm

\headheight 0pt

\topskip 0mm
\addtolength{\baselineskip}{0.20\baselineskip}
\begin{flushright}
hep-th/9910142
\end{flushright}
\vspace{0.5cm}
\begin{center}
    {\large \bf  Use of Physical Variables in the Chern-Simons Theories}
\end{center}
\vspace{0.5cm}

\begin{center}
 Mu-In Park
\footnote{Electronic address: mipark@physics3.sogang.ac.kr}
\footnote{Talk given at {\it Workshop on Physical Variables in Gauge
    Theories}, Sept. 21-25, 1999, Dubna, Russia.}
\\{\it Department of Physics, Yonsei University, \\
Seoul 120-749, Korea}
\end{center}
\vspace{0.5cm}
\begin{center}
    {\bf ABSTRACT}
\end{center}
The use of the physical variables in the fashion of Dirac in the
three-dimensional Chern-Simons theories is presented. Our previous
results are reinterpreted in a new aspect.  
\vspace{3cm}
\vspace{2.5cm}
\begin{flushleft}
\end{flushleft}

\newpage
{\bf I. INTRODUCTION}\\

This talk is devoted to the use of the physical variables in the
fashion of Dirac \cite{Dir:55} to the Chern-Simons (CS) theories in 2+1
dimensions with {\it reinterpretations} of our previous
results \cite{Park:98}. In order to give motivations of our work, it would be better to introduce
the CS theories \cite{Des:82} to this audience first. 

The (pure) CS Lagrangian is defined by 
\begin{eqnarray}
{\cal L}_{CS}=\frac{\kappa}{2} \epsilon^{\mu \nu \rho}
A_{\mu}\partial_{\nu}A_{\rho},
\end{eqnarray}
where $\epsilon^{012}=1$, $g_{\mu \nu}$=diag(1,--1,--1). But this
theory is not so interesting in the aspect of the physical
variables which are gauge invariant because there is no non-vanishing
gauge invariant variable according to the equations of motion,
$F_{\mu \nu} =0$; only the pure-gauge variables are remained. Because
of this problem, let us now consider the CS theories with matter
couplings. Here, we only consider the coupling to the charged
(complex) scalar fields for simplicity (the coupling with fermion
fields is not different much) \cite{Hag:84,Pol:88,Sem:88}: 
\begin{eqnarray}
{\cal L}=\frac{\kappa}{2} \epsilon^{\mu \nu \rho}A_{\mu}\partial_{\nu}A_{\rho}
+(D_{\mu}\phi)^{*}(D^{\mu}\phi)-m^{2} \phi^{*} \phi,
\end{eqnarray}
where $D_{\mu}=\partial_{\mu}+iA_{\mu}$. Then, ${\cal L}$ is invariant up to the total divergence 
under the gauge transformations
\begin{eqnarray}
\label{eq:gaugetransf} 
\phi \rightarrow e^{-i\Lambda } \phi,~~A_{\mu} \rightarrow A_{\mu} +
\partial_{\mu}\Lambda. 
\end{eqnarray}
The equations of motion are given by 
\begin{eqnarray}
\frac{\kappa}{2} \epsilon^{\mu \alpha \beta} F_{\alpha \beta}=J^{\mu}
\end{eqnarray}
with the conserved current $J^{\mu}=i [ (D^{\mu} \phi)^{*} \phi
-\phi^{*} D^{\mu} \phi]$; in the component form they are expressed by
\begin{eqnarray}
\label{eq:B}
&&\kappa B =J^0,\\
\label{eq:E}
&&\kappa E^k =\epsilon_{ki} J^i
\end{eqnarray}
with the (scalar) magnetic and electric fields, $B=\frac{1}{2}
\epsilon^{ij} F_{ij}, E^k =F^{k0}$, respectively. Here, now one can find
the non-vanishing gauge invariant variables and hence this theory is more interesting in the
aspect of the physical variables. Moreover, the integral form of the
constraint (\ref{eq:B}), which is given by 
\begin{eqnarray}
\Phi = \int B~ d^2 {\bf x}  = \frac{1}{\kappa} Q,
\end{eqnarray}
implies that when there is a particle with charge $Q$, it carries
 the magnetic flux $\Phi=\frac{1}{\kappa} Q$. In this sense, the
CS theory realizes the Wilczek's charge-flux composite \cite{Wil:82}
and this is a role of the CS field to the matters in the particle
 picture \cite{Jac:90}. Its (quantum) field theoretical analysis was first
 given by Hagen in the Coulomb gauge, i.e., $\nabla \cdot {\bf A}
 =0$ \cite{Hag:84}. The (formal) solution of gauge fields $A^{\mu}$ which
 solves the constraints (\ref{eq:B}), (\ref{eq:E}) are given by
\begin{eqnarray}
&&A^i ({\bf x}, t) =\frac{1}{2 \pi \kappa} \int d^2 {\bf y}~ \epsilon_{ij}
\frac{(x-y)^j}{|{\bf x}-{\bf y}|^2} J^0 ({\bf y}, t), \\
&&A^0 ({\bf x}, t) =\frac{1}{2 \pi \kappa} \int d^2 {\bf y}~ \epsilon_{ij}
\frac{(x-y)^j}{|{\bf x}-{\bf y}|^2} J^i ({\bf y}, t). \nonumber
\end{eqnarray}  
In this solution, we obtain first of all the following commutation
relations,
\begin{eqnarray}
&& [Q, \phi ({\bf x}, t) ]=  \phi({\bf x},t), \\
&& [B({\bf y}, t), \phi ({\bf x},t) ] =\frac{1}{\kappa} \delta^2 ({\bf x}-{\bf
  y}) \phi ({\bf y}, t),   
\end{eqnarray}
where $Q=\int J^0 d^2 {\bf x} =i \int  d^2 {\bf x}~(\pi \phi-\phi^* \pi^*)$.
These represent that field $\phi (x)$ carries the unit charge and
delta-function magnetic field, $\frac{1}{\kappa} \delta^2 ({\bf
  x}-{\bf y})$ at $x$ point; these describe the charge-flux
composite in the context of quantum field theory. Moreover, if we
consider the angular momentum generator which is constructed from the
{\bf symmetric} gauge invariant energy-momentum tensor, we can find there
is unconventional term in addition to the expected orbital angular momentum
part:
\begin{eqnarray}
M^{12} &=&\int d^2 {\bf x}~\epsilon_{ij} x^i T^{0j} \nonumber\\
       &=& \int d^2 {\bf x}~\epsilon_{ij} x^i (\pi D^j \phi + c.c ) \nonumber\\ 
       &=&\int d^2 {\bf x}~ (\pi {\bf x} \times \nabla \phi +c.c.) + \frac{1}{2
       \pi \kappa} \int d^2 {\bf x}~ \int d^2 {\bf y}~ J^0 (x) \frac{{\bf x} \cdot ( {\bf
       x}-{\bf y})} {|{\bf x}-{\bf y}|^2} J^0 (y) \nonumber\\
       &=& J_{orb} + \frac{1}{4 \pi \kappa} Q^2.
\end{eqnarray}
The first term is the usual orbital angular momentum part. The second
term is an unconventional term which can not be removed by a redefinition
of the angular momentum generator because this anomalous term is uniquely
governed by $[ M^{01}, M^{02} ]=i M^{12}$ in the
Poincar\'e algebra \cite{Hag:84,Hag:85}. Now, it is easy to find that the first and second
terms produce the following transformation for $\phi$: 
\begin{eqnarray}
[ M^{12}, \phi ] =-i {\bf x} \times \nabla \phi + \frac{1}{2 \pi
  \kappa} Q \phi.
\end{eqnarray} 
Here, the first term shows the orbital angular momentum and the second one
shows the (anomalous) {\it spin for scalar fields} which is a very surprising
result. [This unconventional transformation is usually called as the {\it rotational anomaly} but here
the use of word `{anomaly}' is not usual one because the anomaly 
exists even at the classical label in our case.] This is the first non-trivial result in the theory which was
given first by Hagen \cite{Hag:84,Hag:85}. In addition to this, several authors considered
furthermore the so-called {\it anyonic} [here, `{anyon}' means the
object which interpolates between boson and fermion] commutation relation inspired
by the works of Polyakov \cite{Pol:88} and Semenoff \cite{Sem:88}: In
the Semenoff's work, for example, he introduced the composite
operators like as
\begin{eqnarray}
\label{eq:Semenoff}
&&\bar{\phi} ({\bf x}) := exp\left[\frac{i}{2 \pi \kappa} \int d^2 {\bf
  y}~ \Theta ({\bf x}-{\bf y})
  J^0 ({\bf y}) \right] \phi ({\bf x}), \\
&&\bar{\pi} ({\bf x}) := exp\left[-\frac{i}{2 \pi \kappa} \int d^2 {\bf
  y}~ \Theta ({\bf x}-{\bf y})
  J^0 ({\bf y}) \right] \pi (x)  \nonumber 
\end{eqnarray}
and he found that their commutation relations are given by
\begin{eqnarray}
&&\bar{\phi} ({\bf x}) \bar{\phi} ({\bf y}) -e^{-\frac{i \Delta}{2 \pi \kappa} }
 \bar{\phi} ({\bf y}) \bar{\phi} ({\bf x}) =0, \\ 
&&\bar{\pi} ({\bf x}) \bar{\pi} ({\bf y}) -e^{\frac{i \Delta}{2 \pi \kappa} }
 \bar{\pi} ({\bf y}) \bar{\pi} ({\bf x}) =0, \nonumber\\ 
&&\bar{\phi} ({\bf x}) \bar{\pi} ({\bf y}) -e^{\frac{i \Delta}{2 \pi \kappa} }
 \bar{\pi} ({\bf y}) \bar{\phi} ({\bf x}) =i \delta^2 ({\bf x}-{\bf
 y}) \nonumber,
\end{eqnarray}
where, $\Theta ({\bf x}-{\bf y})$ is the angle between the vector ${\bf x}-{\bf y}$ and
the $x_1$-axis and $\Delta:= \Theta({\bf x}-{\bf y}) -\Theta({\bf
  y}-{\bf x}) =\pi (mod~2
\pi)$. So, depending on the value of the CS coupling constant
$\kappa$, the composite operators can represent the fermions
($\frac{\Delta}{2 \pi \kappa}=\pi$) or
generally anyons ($\frac{\Delta}{2 \pi \kappa}\neq \pi$)as well as
bosons ($\frac{\Delta}{2 \pi \kappa}=0$). These are all stories about
the CS theories in the canonical quantization with the Coulomb gauge. Now,
our question is \\ 

{\it What is the physical (gauge independent) effect and
  what is the unphysical (gauge artifact) one ? } \\

{\bf II.  GAUGE INVARIANT FORMULATION}\\

In order to shed some light to the (gauge artifact) problem, let us
consider the Dirac's physical variables instead of gauge
varying base fields $\phi, \pi,\cdots, etc.$ \cite{Dir:55,Park:98}
\begin{eqnarray}
\label{eq:phy.var.}
\hat{\phi}({\bf x}) := \phi({\bf x}) e^{iW},~
\hat{\pi}({\bf x}) := \pi({\bf x}) e^{-iW},~
{\cal A}_{\mu}({\bf x}):={A}_{\mu}({\bf x})-\partial_{\mu}W,
\end{eqnarray}
and their complex conjugates with $W=\int d^{2} {\bf z}~c_{k}({\bf x},
 {\bf z}) A^{k}({\bf z}) $. These are gauge invariant for {\bf  each } 
Dirac dressing function $c_{k}({\bf x},{\bf z})$ under the gauge
 transformation (\ref{eq:gaugetransf}) if the dressing function  satisfies
\begin{eqnarray}
\partial^{k}_{z} c_{k}({\bf x},{\bf z})=-\delta^2({\bf x}-{\bf z}). 
\end{eqnarray}
Then, it is easy to find that the physical variable
    $\hat{\phi}({\bf x})$, which is gauge invariant \footnote{Here, one should distinguish {\it gauge
    invariance} and {\it gauge independence}: Our physical variables
  are gauge invariant but they are gauge {\it dependent} because its
  explicit form are gauge dependent. I'd like to thank 
  Dr. J. Watson for providing a source of this note in his talk \cite{J:99}.},
carries one unit charge
$(~[Q ({\bf y}), \hat{\phi}({\bf x}) ]= \hat{\phi}({\bf x})
~)$ and point magnetic field $([B ({\bf y}), \hat{\phi}({\bf x}) ]=\frac{1}{\kappa}
\delta^2 ({\bf x}-{\bf y}) \hat{\phi}({\bf x}) )$ at the matter point
together with the vector potential around the matter point $([{A^i} ({\bf y}), \hat{\phi}({\bf x}) ]=\frac{1}{\kappa}
\epsilon_{ik} c_{k} ({\bf x}, {\bf y}) \hat{\phi}({\bf x}))$.
This situation corresponds to the dual picture of the QED case where 
$\hat{\phi}({\bf x})$ 
carries the radial ${\bf E}$ field together with the scalar potential
around the matter point \cite{Dir:55,Park:98b}. 

Next, let us consider the space-time transformation properties of the
physical variables. To this end, let us consider the improved Poincar\'e 
generators \cite{Cal:70} constructed from the symmetric (Belinfante)
energy-momentum tensor \cite{Bel:40} which being (manifestly) gauge
invariant and satisfying the Poincar\'e algebra:
\begin{eqnarray}
&&{P}^{0} = \int d^{2} {\bf x}\left[ |{\pi}|^{2}
      +|{D}^{i} {{\phi} } |^{2} +m^{2} |{{\phi}}|^{2}
       \right], \nonumber \\
&&{P}^{i} = \int d^{2}{\bf x} \left[{{ \pi}}{D}^{i}{{\phi}} 
      +({D}^{i}{{\phi}})^{*} {{\pi}^{*}}
       \right], \nonumber \\
&&{M}^{12} = \int d^{2}{\bf x}~ \epsilon_{ij}x^{i} \left[
  {{ \pi}}{ D}^{j}{{\phi}} 
  +({D}^{j} {{\phi}})^{*}  {{\pi}^{*}}
   \right], \nonumber \\
&&{M}^{0i} = x^{0} {P}^{i}-
 \int d^{2}{\bf x}~ x^{i}\left[|{{ \pi}}|^{2} 
 +|{D}^{j}{{\phi}}|^{2} +m^{2} |{{\phi}}|^2 \right].
\end{eqnarray}
First of all, we consider the 
spatial translation generated by
\begin{eqnarray}
[ {\hat{\phi}}({\bf x}),{P}^{j}]=
\partial^j {\hat{\phi}}({\bf x}) -i {\hat{\phi}}({\bf x})
\int d^2 {\bf z} \left(\partial^j_z c_k ({\bf x},{\bf z}) + \partial^j_x c_k
({\bf x},{\bf z})\right) {A}^k ({\bf z}).
\end{eqnarray}
This shows the translational anomaly in general. However, we assume 
that this anomaly should not appear in order that  
 ${\hat{\phi}}$ responds conventionally to translations
because the translation invariance is a genuine property of
space-times which is independent on the matter contents. With 
this assumption, we obtain the condition that
$c_{k}({\bf x}, {\bf z})$ be translationally invariant
\begin{eqnarray}
\partial^i_z c_k ({\bf x},{\bf z})=-\partial^i_x c_k ({\bf x},{\bf z}),
\end{eqnarray} 
i.e., $c_k({\bf x},{\bf z})=c_k({\bf x}-{\bf z})$.
Furthermore, this condition also guarantees the correct spatial
translation law for another physical field in (\ref{eq:phy.var.}):
$[{\cal A}_i( {\bf x}), P^j ] =i \partial ^j {\cal A}_i ({\bf x})$.

By applying similar assumption to the time translation we obtain
\begin{eqnarray}
{[ {\hat{\phi}}({\bf x}),P^0 ]}&=&
i \partial^0 {\hat{\phi}}({\bf x}) + {\hat{\phi}}({\bf x})
\int d^2 {\bf z}~ \partial^0 ( c_k ({\bf x}-{\bf z}) ) A^k ({\bf z})\nonumber \\
&:=&i \partial^0 {\hat{\phi}}({\bf x}), \\
{[ {\cal A}_i ( {\bf x}), P^0 ]} &=&i \partial^0 {\cal A}_i ({\bf x}) \nonumber
\end{eqnarray}
with a necessary condition of
\begin{eqnarray}
\partial^0 c_k ({\bf x}-{\bf z})=0.
\end{eqnarray} 

However, for the rotation and Lorentz boost, the anomaly is present as
the spin or other properties of $\hat{\phi}, {\cal A}_i $:
\begin{eqnarray}
     &&[ \hat{\phi}({\bf x}), M^{12} ]
= i\epsilon_{ij} x_i \partial_j \hat{\phi}({\bf x}) +
       \Xi^{12}({\bf x}) \hat{\phi}({\bf x}),  \nonumber\\
&&[ {\cal A}_i ({\bf x}), M^{12}]
= i \epsilon_{jk} x_j \partial_k  {\cal A}_i ({\bf x}) -i\epsilon_{ij}
 {\cal A}_j ({\bf x})+ i\partial_i \Xi ^{12}({\bf x}), \nonumber\\
 &&[ \hat{\phi}({\bf x}), M^{0j} ]
= i x^0 \partial^j \hat{\phi}({\bf x})  
 -i x^j \partial^0 \hat{\phi} ({\bf x}) + \Xi ^{0j}({\bf x})\hat{\phi}
 ({\bf x}), \nonumber \\ 
&&[ {\cal A}_i({\bf x}), M^{0j} ]
 =  ix^0 \partial^j {\cal A}_i({\bf x})  -i x^j \partial^0 {\cal
  A}_i({\bf x})-i\delta_{ij} {\cal A}_0 + i\partial_i  \Xi 
 ^{0j}({\bf x}) \phi ({\bf x});  
\end{eqnarray}
or in the compact form these are expressed as follows
\begin{eqnarray}
\label{eq:ano_transf}
[ {\cal F}_{\alpha}({\bf x}), M^{\mu \nu} ]&=& 
   i x^{\mu} \partial^{\nu}{\cal F}_{\alpha}({\bf x})
   -i x^{\nu} \partial^{\mu} {\cal F}_{\alpha}({\bf x}) 
   +i \Sigma^{\mu \nu}_{\alpha \beta} {\cal F}_{\beta}({\bf x})
   +i \Omega^{\mu \nu}_{\alpha}({\bf x}), \nonumber \\
\Omega^{\mu \nu}_{\hat{\phi} }({\bf x})&=&
   -i {\Xi}^{\mu \nu} ({\bf x}) 
\hat{\phi} ({\bf x}), ~\Omega^{\mu \nu}_{ \hat{\phi}^{*} }({\bf x})=
   i {\Xi}^{\mu \nu} ({\bf x}) 
\hat{\phi}^* ({\bf x}), \nonumber \\
\Omega^{\mu \nu}_{{\cal A}_i}({\bf x})&=&
  \partial_i {\Xi}^{\mu \nu} ({\bf x}),
\end{eqnarray}
where
\begin{eqnarray}
&&\Xi ^{\mu \nu} =- \Xi^{\nu \mu}, \nonumber \\
&&{\Xi}^{12}({\bf x})= -{\bf x} \times {\bf \cal A} ({\bf x})
+ \frac{1}{\kappa}
 \int d^2 {\bf z} ~ {\bf z}\cdot {\bf c}({\bf x}-{\bf z}) J_0({\bf z}),\nonumber \\
&&{\Xi}^{0i}({\bf x})=-x_i {\cal A}^0({\bf x})
+\frac{1}{\kappa} \int d^2 {\bf z} ~ z_i~ 
  {\bf  c}({\bf x}-{\bf z}) \times {\bf J}({\bf z})
\end{eqnarray}
with  $ {{\cal  F}}_{\alpha}= ({{\cal A}_{\mu}},
{\hat{\phi}},{{\hat{\phi}}^*})$ and
the spin-factors $\Sigma ^{\mu \nu}_{\alpha \beta}= \eta ^{\mu
  \alpha} \eta ^{\nu}_{\beta} -\eta^{\mu}_{\beta} \eta^{\nu \alpha},
  ~\Sigma ^{\mu \nu}_{\phi (\phi^*)}=0 $
for the gauge and scalar fields, respectively.
The $\Omega^{\mu \nu}_{\alpha}$ represents the gauge {\it invariant}
  anomaly  term. Now, I would like to note several interesting aspects
of our formulation in the followings.\\

{\bf A. On-shell expression of ${\cal A}_{\mu}$}\\

The physical
variable ${\cal A}_{\mu}$ in (\ref{eq:phy.var.}) is rather formal
one. But, if we use the equations
of motion (\ref{eq:B}), (\ref{eq:E}), we can obtain the following expressions:
\begin{eqnarray}
{\cal A}_i ({\bf x}) &\approx& -\frac{1}{\kappa} \int d^2 {\bf z}~
\epsilon_{ik} c_{k} ({\bf x}-{\bf z}) J^0 ({\bf z}), \nonumber \\
{\cal A}_0 ({\bf x}) &=& \frac{1}{\kappa} \int d^2 {\bf z}~
 {\bf c} ({\bf x}-{\bf z}) \times {\bf J} ({\bf z}), 
\end{eqnarray}
respectively. Then, it is straightforward to show that the anomaly
terms can be written in the simple forms:
\begin{eqnarray}
\Xi^{12} ({\bf x}) &\approx& -\frac{1}{\kappa} \int d^2 {\bf z}~
({\bf x}-{\bf z}) \cdot {\bf  c} ({\bf x}-{\bf z}) J^0 ({\bf z}), \nonumber \\
\Xi^{0i}({\bf x}) &=& \frac{1}{\kappa} \int d^2 {\bf z}~
 (x^i -z^i)~ {\bf c} ({\bf x}-{\bf z}) \times {\bf J} ({\bf z}). 
\end{eqnarray}
This can be considered as a {\it generalization of the Coulomb gauge 
to the arbitrary $c_{k}$ case}: With a particular choice of $c_k$ which
corresponds to the Coulomb gauge, this reduces to the previous result in
Coulomb gauge. \\

{\bf B. Gauge fixing vs ``$c_k$''}\\

Now, I'd like to note an issue, which has been controversial, about the
connection of gauge fixing and ``$c_k$'' function. The problem which
makes the situation complicate is that the connection is {\bf not} {\it one
to one} in general; in most case of (partial) gauge fixings which
allow some remaining gauge transformations, $c_k$ is not uniquely
determined. But there is one exceptional case of the Coulomb gauge where the
connection is one to one and so only one $c_k$ is allowed.\footnote{However, 
when the gauge fixing is not introduced, the fixing $c_k$ does not mean 
fixing a gauge.} 
Let us consider these two cases in detail. \\

{\bf  a) axial gauge, $A_1 \approx 0$}\\

 In this case, the gauge
transformation function $\Lambda$ should be $x^1$ independent, i.e.,
$\partial_1 \Lambda =0$ in order to preserve the given gauge choice under
the gauge transformation: $A_1 \rightarrow A_1 +\partial_1 \Lambda :=0,~
A_2 \rightarrow A_2 +\partial_2 \Lambda $. On the other hand, the physical
matter field $\hat{\phi}$, which is gauge invariant,  becomes 
\begin{eqnarray*}
\hat{\phi}=e^{i \int c_k A^k } \phi =e^{i \int c_2 A^2 } \phi
\end{eqnarray*} 
in this gauge. So, in order to make it gauge invariant in that gauge, 
\begin{eqnarray*}
\hat{\phi} \rightarrow e^{i \int c_2 (A^2 -\partial^2_z \Lambda )}
e^{-i \Lambda} \phi := \hat{\phi} 
\end{eqnarray*}
one should require
\begin{eqnarray}
\partial^2_z c_2 ({\bf x}-{\bf z}) =-\delta^2 ({\bf x}-{\bf z}),~
\partial^1_z c_1 ({\bf x}-{\bf z}) =0.
\end{eqnarray}
The general solution of this equation can be expressed by two
arbitrary functions, $F$ and $G$:
\begin{eqnarray}
&&c_1 ({\bf x}-{\bf z}) =F(x_1 -z_1), \nonumber \\
&&c_2({\bf x}-{\bf z}) =-\delta (x_1-z_1) \epsilon (x_2 -z_2) + G(x_1 -z_1).
\end{eqnarray}
The existence of these two arbitrary functions makes the analysis complicate. But for the
simplest case of $F=G=0$,  the physical matter field $\hat{\phi}$
becomes
$\hat{\phi} ({\bf x}) =e^{i \int ^{x_2}_{-\infty} dz_2 A^2 (x_1, z_2
)} \phi ({\bf x})$ and it carries the vector potential tail along
straight line from the matter point, $A^1 ({\bf y})=-\frac{1}{\kappa}
\delta (x_1 -y_1) \epsilon (x_2 -y _2)$; this is the usual
line-integral form of the physical variables \cite{Dir:55}. \\

{\bf  b) Coulomb gauge, $\nabla \cdot {\bf A}=0$ }\\

 In this case, the gauge
transformation function $\Lambda$ should satisfy Laplace equation $\nabla^2
\Lambda =0$ over {\it all} space in order to maintain the given gauge under
the gauge transformation: $\nabla \cdot {\bf A} \rightarrow
\nabla \cdot {\bf A}+ \nabla ^2 \Lambda :=0$. However, since there is only the
trivial solution of $\Lambda =0$ for the Laplace equation over all space, we don't
need to introduce the gauge compensating factor in the physical
matter field: $\hat{\phi} =\phi$. In other words, $W=\int d^2 {\bf z}~ c_j ({\bf
x},{\bf z}) A^j ({\bf z}):=0$ in the general formula
(\ref{eq:phy.var.}). Then, it is to not difficult to show that there
is a unique solution of 
\begin{eqnarray}
c^j({\bf x}-{\bf z}) =-\frac{1}{2 {\pi}}
 \frac{({\bf x}-{\bf z})^j}{|{\bf x}-{\bf z}|^2}.
\end{eqnarray}
Hence, this case corresponds to a {\it complete} gauge fixing.\footnote{A generalization of this case was known \cite{Gir:90}.}\\

{\bf C. (Anyonic) Commutation relations}\\

The next thing I would like to note is the commutation relations of the physical
variables which show the anyonic commutation relations in general:
\begin{eqnarray}
\label{eq:graded_comm}
&&\hat{\phi}({\bf x})\hat{\phi}({\bf y})   
-e^{- \frac{i}{\kappa} \Delta ({\bf x}-{\bf y})} 
  \hat{\phi}({\bf y})\hat{\phi}({\bf x}) =0,\nonumber \\
&&\hat{\phi}({\bf x})\hat{\pi}({\bf y})- 
e^{\frac{i}{\kappa}\Delta ({\bf x}-{\bf y})} 
\hat{\pi}({\bf y})\hat{\phi}({\bf x}) = \delta^2({\bf x}-{\bf y}), \nonumber \\ 
&&\left[ {{\cal A}_i}({\bf x}), \hat{\phi}({\bf y}) \right]
  =-\frac{1}{\kappa} \hat{\phi}({\bf y}) \left
  [ \epsilon_{ik}c_k({\bf y}- 
  {\bf x})
   + \partial^x_i \Delta({\bf x}-{\bf y}) \right], \nonumber \\ 
&&\left[ {{\cal A}_i}({\bf x}),{{\cal A}_j}({\bf y}) \right] =
  \frac{i }{\kappa} \left[\epsilon_{ij}\delta^2({\bf x}-{\bf y})
  +\xi_{ij}({\bf x}-{\bf y}) + \partial_i^x \partial_j^y 
\Delta ({\bf x}-{\bf y}) \right],~ \cdots,~ etc.,
\end{eqnarray}
where we have introduced two functions
\begin{eqnarray}
\label{eq:Delta}
&&\Delta({\bf x}-{\bf y})=\int d^2 {\bf z}~ \epsilon^{kl} c_k({\bf x}-{\bf
  z}) c_l({\bf y}-{\bf z}) , \nonumber \\
\label{eq:Xi}
&&\xi_{ij}({\bf x}-{\bf
  y})=\epsilon_{ik}\partial^y_j c_k({\bf y}-{\bf x}) -
\epsilon_{jk}\partial^x_i c_k({\bf x}-{\bf y}),
\end{eqnarray} 
which are totally antisymmetric under the interchange
of all the indices. If we consider the particular case of Coulomb
gauge, we can find that $\Delta({\bf x}-{\bf y})=0~ [\xi_{ij}({\bf x}-{\bf
  y})=-\epsilon_{ij} \delta^2 ({\bf x}-{\bf y}) ]$. But in general
this is not true ! This implies that the physical variable
$\hat{\phi}$ satisfies the {\it anyonic commutation relation 
depending on $c_{k}$}: Hence, this can be a {\it natural } definition of
the anyon in the {\bf minimal form} of (\ref{eq:phy.var.}) !; we can
introduce any arbitrary gauge invariant factors except for the minimally
introduced gauge compensating factor of (\ref{eq:phy.var.}) but {\it there
is no unique way to fix them}.\footnote{In the particular case of
non-relativistic delta-function sources, there is an unique way to
define the additional factor \cite{Hag:89}. Moreover, there is an
interesting formulation where the additional factor is determined by
the temporal non-locality \cite{Lav:93} contrast to ours where the
temporal {\it locality} was assumed from the start in (\ref{eq:phy.var.}).
But it is unclear that their formulation provides a new
determination of the factor since there is no retardation
effect; there are only the instantaneous interactions in our case.} 
On the other hand, we note that the Semenoff's construction of (\ref{eq:Semenoff}), 
what we have introduced in
the former part of this talk, can be considered like as \cite{Lee:92}
\begin{eqnarray}
\bar{\phi} ({\bf x}) = exp \left[\frac{i}{2 \pi \kappa} \int d^2 {\bf
  y}~ \Theta ({\bf x}-{\bf y})
  J^0 ({\bf y}) \right] \hat{\phi} ({\bf x})
\end{eqnarray}
in our formulation; these operator has the (gauge invariant)
exchanging phase factor of (14) in addition to the phase factor of
(30) for the minimal part $\hat{\phi}$.
\\

{\bf D. Physical and unphysical effects of the rotational anomaly for
$\phi$}\\

Now, to discuss the gauge artifact problem on the rotational
anomaly let us note the following general relation 
\begin{eqnarray} 
M^{12} \approx M^{12}_c -\frac{\kappa}{2} 
\int d^2 {\bf z}~ \partial ^k (z^k {\bf A} \cdot
{\bf A}-{\bf z}\cdot {\bf A} A^k ),
\end{eqnarray} 
where $M^{12}_c$ is the canonical angular momentum
\begin{eqnarray}
M^{12}_c =\int d^2 {\bf z} \left\{ - {\bf z} \times [\pi \nabla \phi + 
(\nabla \phi)^* \pi^* ] +\kappa {\bf z} \cdot {\bf A} (\nabla \cdot {\bf A}) +\frac{\kappa}{2}
 \partial^k ({\bf z} \cdot {\bf A} A^k)\right\}.
\end{eqnarray}
The surface terms in $M^{12} -M^{12}_c$ and $M^{12}_c$, which are gauge invariant for the rapidly 
decreasing gauge transformation function $\Lambda$, give the gauge independent spin terms 
``$({1}/{4 \pi \kappa}) Q^2$" (unconventional) and ``$0$" (conventional)
 in $M^{12}$, respectively. 
From which the anomalous spin $\frac{Q}{2 \pi \kappa}$ of (12) for the matter field is 
readily seen to follow for general gauges.
On the other hand, the second term of $M^{12}_c$, which vanishes only in the 
Coulomb gauge, gives for the general gauges 
the gauge restoring contribution to the rotation transformation for the matter
field. This can be summarized by
\begin{eqnarray}
[ M^{12}, \phi ] =-i {\bf x} \times \nabla \phi + \frac{1}{2 \pi
  \kappa} Q \phi + (\mbox{gauge restoring terms for non-rotation
  invariant gauge}),
\end{eqnarray} 
in which the second term is the {\it gauge invariant}, {i.e.,}
physical (anomalous) spin term and the last one is gauge dependent
artifact term.\\

{\bf III. CONCLUSION}\\

In summary, we have shown that the use of the Dirac's physical
variables in the CS-matter theories are: \\
a) generalization of the Coulomb gauge result to arbitrary $c_k$
cases, \\
b) providing a natural definition of anyons modulus the gauge
invariant factors {\it which will be determined by more fundamental
theorems}.

Involving the gauge artifact problems, what have been clarified in this talk are: \\
a) the anomalous spin of the base matter field is a physical effect
and not the gauge artifact; and hence the Coulomb gauge is the best one
to obtain the physical spin directly. On the other hand, since the boson 
commutation relation of the base matter field, $[\phi ({\bf x}), \phi({\bf y}) ]=0$ 
is gauge independent (thought not gauge invariant), this is a counter 
example \cite{Hag:84} for the {\it quantum mechanical} spin-statistics connection 
\cite{Wil:82}. \\
b) the minimal form of the physical matter fields is anyon field
(i.e., satisfies the anyonic commutation relations) but
depending on ${c_k}$; its physical contents depend on $c_k$ explicitly. 
The gauge artifact nature of this anyon, when a gauge fixing 
is introduced, is not so manifest because of complication in relation of 
${c_k}$ and the gauge fixing. This problem has not been completely resolved yet;
though the physical matter field with additional gauge invariant factor
has the physical exchanging phase factor additionally, the problem for
the minimal part remains. 

As a further work, the non-Abelian generalization will be
interesting. But I have no idea at present mainly because our result 
depends on the detailed expression of the physical variables
\footnote{This is contrast to the formulation of Nair $et. al$
  \cite{Nai:96} where the details of the physical variables are not
  important.} which will be very non-trivial ones \cite{Bel:96}.\\

I would like to thank Prof. Young-Jai Park for providing
the financial support through the Korea Research
Foundation (KRF), Project No. 1998-015-D00074 and organizers of this 
workshop and other members in the JINR for warm hospitality. The
present work was also supported in part by the Korea Science and Engineering 
Foundation (KOSEF), Project No. 97-07-02-02-01-3, in part by KRF under 
project No. 98-015-D00061, and also in part by a 
postdoctoral grant from the Natural Science Research Institute, Yonsei 
University in the year 1999.


\end{document}